# A Strong Bond Model for Stress Relaxation of Soft Solid Interfaces

Arun K. Singh* and Vinay A. Juvekar**
*Department of Mechanical Engineering, VNIT Nagpur, Nagpur-440010, India
**Department of Chemical Engineering, IIT Bombay, Mumbai-400076, India

**Abstract**

In this article, we propose a mathematical model which explains the formation of strong bonds during the relaxation process of a soft solid on a hard surface. As a result, the soft solid relaxes to a non zero residual stress level. The model assumes that formation of strong bonds occurs owing to transition from weak to strong bonds at a critical time. Parametric studies are carried out to understand the effect of different friction parameters related with the model on the relaxation process. The relaxation model is, in turn, validated with experiment and corresponding numerical values are justified.

**Introduction**

Stress relaxation is generally seen in slide-hold-slide (SHS) experiments[1-3]. This phenomenon gives rise to many frictional observations for instance waiting time and shear velocity dependent frictional strength, stick-slip vibration etc[1-4]. Dieterich[2] and Marone[3] have reported stress relaxation in their SHS friction experiments on hard surfaces. The rate and state friction (RSF) model is used to explain the relaxation process on a hard surface[3,5]. Persson also studied the relaxation process theoretically for hard surfaces[4]. Baumberger has reported stress relaxation of a gelatin gel on the glass surface[1]. However, there is no model available in literature for soft solids such as gels and elastomers which can explain its relaxation process to a non zero stress level on a hard surface. We consider the application of the population balance approach to model the stress relaxation process[6].

In the relaxation experiment, a puller, pulling the top face of the soft block under direct shear at a constant velocity $V_0$, is brought to a sudden halt. Since the block is under deformed state, it exerts force on the puller. With the passage of time, this force reduces to a non-zero stress level due to relaxation is known as the residual stress (dead stress). The reason for the relaxation is that the base of the block, which is in contact with the hard surface, slides on it. Simultaneously, a process on interfacial chains occurs in the direction which reduces the deformation of the block is known as creep. The creep arises due to the fact that the chains, which are bonded to the hard surface, are stretched in the direction of the pull. When they detach from the surface, they contract. When they bond to the surface again, the locations of the bonds have shifted in the direction of the strain. Thus with each make-and-break process, the base of the block moves in the direction of the puller. As a result, strain in the block diminishes with time. The velocity of creep, $V_c$, is highest at the beginning when the strain in the block is the highest. It gradually decreases as the strain reduces with time. Creep velocity, $V_c$ is also expected to be smaller for greater strength of the bonds between the polymer chains and the surface, and for greater stiffness of the bulk of the specimen. In the present study, it will be shown that the weak bond relaxation model is a special case of the present model[5].

Let the puller, attached to the top surface of the specimen under direct shear, is moving with a steady velocity $V_0$ at time $t < 0$. At time $t = 0$, the puller is suddenly halted, i.e., its velocity is brought to zero. The block is allowed to relax from its initial deformed state to the final undeformed state with time. It is assumed that the specimen is perfectly elastic and hence no bulk flow occurs during the period of relaxation process. Our aim is to determine, how the stress $\sigma(t)$ exerted by the block on the slider decreases with time.

If, at time $t$, the upper face of the block is displaced relative to the base, by distance $\Delta L_b(t)$, then the shear strain in the block is given by $\gamma(t) = \Delta L_b(t)/h$ where $h$ is the height of the block. If $G$ is the shear modulus of the material of the block, then the stress exerted by the block on the puller is given by the following equation. Time dependent stress is defined by $\sigma(t) = G\gamma(t)$ therefore

$$\sigma(t) = \left(\frac{G}{h}\right)\Delta L_b(t) = K_b \Delta L_b(t) \qquad (1)$$

Here $K_b$ represents the stiffness per unit area of the soft block which is given by $K_b = G/h$. The strain relaxation is related to the creep velocity $V_c(t)$ of the base of the block by the following equation $V_c(t) = -d\Delta L_b(t)/dt$. The corresponding rate of reduction in the force $F$ is obtained from above as

$$\frac{d\sigma(t)}{dt} = K_b \frac{d\Delta L_b(t)}{dt} = -K_b V_c(t) \qquad (2)$$

The equation for the population balance of the bonds between the base of the block and the hard surface can be written from literature[6] as

$$\frac{\partial n(t_a,t)}{\partial t} = -\frac{\partial n(t_a,t)}{\partial t_a} + \frac{1}{\tau}\left[N_0 - N(t)\right]\delta(t_a) - \frac{u}{\tau}e^{\lambda f(t_a)/k_B T}n(t_a,t) \tag{3}$$

where $u = e^{-W/k_B T}$, and $\tau = \tau_0 e^{E/k_B T}$.

To simplify the analysis, we assume that the relaxation occurs at much slower pace compared to the kinetics of bond formation/rupture, so that the bonds are at the equilibrium age distribution at any instant of time. This quasi-steady state approximation allows us to neglect the time dependent term $\partial n(t_a,t)/\partial t$ on the left hand side of Eq 3 and write the equation as

$$\frac{\partial n(t_a,t)}{\partial t_a} = \frac{1}{\tau}\left[N_0 - N(t)\right]\delta(t_a) - \frac{u}{\tau}e^{\lambda f(t_a,t)/k_B T}n(t_a,t) \tag{4}$$

This approximation is valid only if

$$\left|\frac{\partial n(t_a,t)}{\partial t}\right| \ll \left|\frac{\partial n(t_a,t)}{\partial t_a}\right| \tag{5}$$

or when $O(t) \gg O(t_a)$, that is the order of magnitude of the time period of the bulk relaxation of the block is much longer than the average lifetime of a bond. The time dependence of $n(t_a,t)$ appears through $f(t_a,t)$ which changes with time. The distribution of bonds is a function of time, through its dependence on $V_c(t)$.

We assume that two kinds of bonds are formed between the polymer chains and the substrate. Initially the bonds formed are weak. With aging, the bonds become stronger. There are two types of concept to account for the transition from weak to strong bonds. One of these two is discontinuous transition from weak to strong bond and this will be discussed in the next section.

**Discontinuous Transition from Weak to Strong Bond**

At low values of $t_a$, the bond is weak. Beyond a certain value $t_{ws}$ of $t_a$, the bond shifts to a stronger variety. Let the values of $u$, for weak bonds be denoted by $u_w$, and for strong bonds by $u_s$. Note that $u_w \gg u_s$. We can write

$$u = \begin{cases} u_w : t_a < t_{ws} \\ u_s : t_a > t_{ws} \end{cases} \tag{6}$$

Let the corresponding population densities be denoted by $n_w$ and $n_s$. For $t_a < t_{ws}$, the following equation holds

$$\frac{\partial n_w(t_a,t)}{\partial t_a} = -n_w(t_a,t)\frac{u_w}{\tau_w}e^{\lambda_w f(t_a,t)/k_B T} \tag{7}$$

with the initial condition

$$n_w(0,t) = \frac{1}{t_w}(N_0 - N(t)) \tag{8}$$

For $t_a > t_{ws}$, the following equation holds

$$\frac{\partial n_s(t_a,t)}{\partial t_a} = -n_s(t_a,t)\frac{u_s}{\tau_s}e^{\lambda_s f(t_a,t)/k_B T} \tag{9}$$

We assume that all weak bonds are converted into strong bonds at $t = t_{ws}$. This gives us

$$n_w(t_{ws},t) = n_s(t_{ws},t) \tag{10}$$

Integration of Eq7 subject to the initial condition in Eq8 yields, the number density of weak bonds as

$$n_w(t_a,t) = \frac{1}{t_w}(N_0 - N(t))g_w(t_a,t) \qquad t_a < t_{ws} \tag{11}$$

where

$$g_w(t_a,t) = \exp\left(-\frac{u_w}{\tau_w}\int_0^{t_a} e^{\lambda_w f(\xi,t)/k_B T}d\xi\right) \tag{12}$$

Integration of Eq 9 subject to boundary condition in Eq 10 yields

$$n_s(t_a,t) = n_w(t_{ws},t)g_s(t_a,t) \qquad t_a > t_{ws} \tag{13}$$

where

$$g_s(t_a,t) = \exp\left(-\frac{u_s}{\tau_s}\int_{t_{ws}}^{t_a} e^{\lambda_s f(\xi,t)/k_B T}d\xi\right) \tag{14}$$

We can obtain the expression for $n_w(t_{ws},t)$ from Eq 11 as

$$n_w(t_{sw},t) = \frac{1}{t_w}(N_0 - N(t))g_w(t_{ws},t) \tag{15}$$

Combining Eq 13 and 15, we obtain

$$n_s(t_a,t) = \frac{1}{t_w}(N_0 - N(t))g_w(t_{ws},t)g_s(t_a,t) \tag{16}$$

The total number of weak bonds at time $t$, that is, $N_w(t)$, is obtained using Eq 11 as

$$N_w(t) = \int_0^{t_{ws}} n_w(t_a,t)dt_a = \frac{1}{t_w}(N_0 - N(t))\int_0^{t_{ws}} g_w(t_a,t)dt_a \tag{17}$$

The total number of the strong bonds is obtained using Eq 16 as

$$N_s(t) = \int_{t_{ws}}^{t_{max}} n_s(t_a,t) dt_a = \frac{1}{\tau_w}(N_0 - N(t)) g_w(t_{ws},t) \int_{t_{ws}}^{t_{max}} g_s(t_a,t) dt_a \quad (18)$$

Where $t_{max}$ is the upper limit of the bond age.

The total number of bonds $N(t)$ is now obtained as

$$N(t) = N_w(t) + N_s(t) \quad (19)$$

Substituting the expressions for $N_w$ and $N_s$ from Eq 17 and 18 into Eq 19 we get

$$N(t) = \frac{1}{\tau_w}(N_0 - N(t)) \int_0^{t_{ws}} g_w(t_a,t) dt_a + \frac{1}{\tau_w}(N_0 - N(t)) g_w(t_{ws},t) \int_{t_{ws}}^{t_{max}} g_s(t_a,t) dt_a \quad (20)$$

Rearrangement yields

$$N(t) = N_0 \frac{\frac{1}{\tau_w}\left[\int_0^{t_{ws}} g_w(t_a,t) dt_a + g_w(t_{ws},t) \int_{t_{ws}}^{t_{max}} g_s(t_a,t) dt_a\right]}{1 + \frac{1}{\tau_w}\left[\int_0^{t_{ws}} g_w(t_a,t) dt_a + g_w(t_{ws},t) \int_{t_{ws}}^{t_{max}} g_s(t_a,t) dt_a\right]} \quad (21)$$

Elimination of $N(t)$ between Eq 11 and 21 yields the following expression for $n_w(t_a,t)$

$$n_w(t_a,t) = N_0 \frac{\frac{1}{\tau_w} g_w(t_a,t)}{1 + \frac{1}{\tau_w}\left[\int_0^{t_{ws}} g_w(t_a,t) dt_a + g_w(t_{ws},t) \int_{t_{ws}}^{t_{max}} g_s(t_a,t) dt_a\right]} \quad (22)$$

In a similar manner, we can obtain $n_s(t_a,t)$ as

$$n_s(t_a,t) = N_0 \frac{\frac{1}{\tau_w} g_w(t_{ws},t) g_s(t_a,t)}{1 + \frac{1}{\tau_w}\left[\int_0^{t_{ws}} g_w(t_a,t) dt_a + g_w(t_{ws},t) \int_{t_{ws}}^{t_{max}} g_s(t_a,t) dt_a\right]} \quad (23)$$

The friction force per unit area is given by

$$S(t) = \int_0^{t_{ws}} n_w(t_a,t) f(t_a,t) dt_a + \int_{t_{ws}}^{t_{max}} n_s(t_a,t) f(t_a,t) dt_a + gV_c(t) N(t) \quad (24)$$

Substitution of $N(t), n_w(t_a,t)$ and $n_s(t_a,t)$ from Eqs 21 to 23 into Eq 24 gives

$$\sigma(t) = N_0 \frac{\frac{1}{\tau_w}\int_0^{t_{ws}} g_w(t_a,t)f(t_a,t)dt_a + \frac{1}{\tau_w}g_w(t_{ws},t)\int_{t_{ws}}^{t_{max}} g_s(t_a,t)f(t_a,t)dt_a}{1+\frac{1}{\tau_w}\left[\int_0^{t_{ws}} g_w(t_a,t)dt_a + g_w(t_{ws},t)\int_{t_{ws}}^{t_{max}} g_s(t_a,t)dt_a\right]}$$

$$+ \gamma V_c N_0 \frac{\frac{1}{\tau_w}\left[\int_0^{t_{ws}} g_w(t_a,t)dt_a + g_w(t_{ws},t)\int_{t_{ws}}^{t_{max}} g_s(t_a,t)dt_a\right]}{1+\frac{1}{\tau_w}\left[\int_0^{t_{ws}} g_w(t_a,t)dt_a + g_w(t_{ws},t)\int_{t_{ws}}^{t_{max}} g_s(t_a,t)dt_a\right]}$$

(25)

**Steady state solution**

At steady state, Eq 25 reduces to

$$\sigma_0 = N_0 \frac{\frac{1}{\tau_w}\int_0^{t_{ws}} g_w(t_a)f(t_a)dt_a + \frac{1}{\tau_w}g_w(t_{ws})\int_{t_{ws}}^{t_{max}} g_s(t_a)f(t_a)dt_a + \gamma V_c \left[\begin{array}{l}\frac{1}{\tau_w}\int_0^{t_{ws}} g_w(t_a)dt_a \\ + \frac{1}{\tau_w}g_w(t_{ws})\int_{t_{ws}}^{t_{max}} g_s(t_a)dt_a\end{array}\right]}{1+\frac{1}{\tau_w}\left[\int_0^{t_{ws}} g_w(t_a)dt_a + g_w(t_{ws})\int_{t_{ws}}^{t_{max}} g_s(t_a)dt_a\right]} \quad (26)$$

We use linear model $f(\Delta L) = M\Delta L - \gamma V_0$ for polymer chains at the interface and note that $\Delta L = V_0 t_a$. Using these expressions in Eq 12 and 14, we get

$$g_w(t_a) = \exp\left(-\frac{u_w k_B T}{\lambda_w \tau_w M V_0}\exp(-\frac{\gamma V_0 \lambda_w}{k_B T})\left\{\exp\left(\frac{\lambda_w}{k_B T}MV_0 t_a\right)-1\right\}\right) \quad (27)$$

$$g_s(t_a) = \exp\left(-\frac{u_s k_B T}{\lambda_s \tau_s M V_0}\exp\left(\frac{-\gamma V_0 \lambda_w \beta \tau_{ws}}{k_B T}\right)\left\{\exp\left(\frac{\lambda_s}{k_B T}MV_0 t_a\right)-\exp\left(\frac{\lambda_s}{k_B T}MV_0 t_{ws}\right)\right\}\right)$$

(28)

We evaluate various integrals as shown in the appendix-I

$$\frac{1}{\tau_w}\int_0^{t_{ws}} g_w(t_a)dt_a = \frac{k_B T}{\tau_w \lambda_w M V_0}\exp\left(\frac{u_w k_B T}{\lambda_w \tau_w M V_0}\exp\left(\frac{-\gamma V_0 \lambda_w}{k_B T}\right)\right)\left\{\begin{array}{l}E_1\left(\frac{u_w k_B T}{\lambda_w \tau_w M V_0}\exp\left(\frac{-\gamma V_0 \lambda_w}{k_B T}\right)\right) \\ -E_1\left(\begin{array}{l}\frac{u_w k_B T}{\lambda_w \tau_w M V_0}\exp\left(\frac{-\gamma V_0 \lambda_w}{k_B T}\right) \\ \exp\left(\frac{\lambda_w}{k_B T}MV_0 t_{ws}\right)\end{array}\right)\end{array}\right\}$$

(29)

where

$$E_1(x) = \int_x^\infty \frac{e^{-y}}{y} dy \qquad (30)$$

Similarly

$$\frac{1}{\tau_w} \int_{t_{ws}}^\infty g_s(t_a) dt_a = \frac{k_B T \tau_s}{\lambda_w \tau_w M \beta V_0} \exp\left[\frac{u_s k_B T}{\lambda_s \tau_s M V_0} \exp\left(\frac{-\gamma V_0 \lambda_w \beta \tau_{ws}}{k_B T}\right) \exp\left(\frac{\lambda_s}{k_B T} M V_0 t_{ws}\right)\right]$$
$$E_1\left[\frac{u_s k_B T}{\lambda_s \tau_s M V_0} \exp\left(\frac{\lambda_s}{k_B T} M V_0 t_{ws}\right)\right] \qquad (31)$$

Next we have,

$$\frac{1}{\tau_w} \int_0^{t_{ws}} g_w(t_a) f(t_a) dt_a$$

$$= \left(\frac{(k_B T)^2}{\lambda_w^2 \tau_w M V_0}\right) \exp\left(\frac{u_w k_B T}{\lambda_w \tau_w M V_0} \exp\left(\frac{-\gamma V_0 \lambda_w}{k_B T}\right)\right) \begin{bmatrix} G_1\left(\frac{u_w k_B T}{\lambda_w \tau_w M V_0} \exp\left(\frac{-\gamma V_0 \lambda_w}{k_B T}\right)\right) - \\ G_1\left(\frac{u_w k_B T}{\lambda_w \tau_w M V_0} \exp\left(\frac{-\gamma V_0 \lambda_w}{k_B T}\right) \exp\left(\frac{\lambda_w}{k_B T} M V_0 t_{ws}\right)\right) \\ -\ln\left(\frac{u_w k_B T}{\lambda_w \tau_w M V_0}\right) \begin{Bmatrix} E_1\left(\frac{u_w k_B T}{\lambda_w \tau_w M V_0} \exp\left(\frac{-\gamma V_0 \lambda_w}{k_B T}\right)\right) \\ -E_1\left(\frac{u_w k_B T}{\lambda_w \tau_w M V_0} \exp\left(\frac{-\gamma V_0 \lambda_w}{k_B T}\right) \\ \exp\left(\frac{\lambda_w}{k_B T} M V_0 t_{ws}\right) \end{pmatrix} \end{Bmatrix} \end{bmatrix}$$

$$(32)$$

The integral $G_1(x)$ is defined as

$$G_1(x) = \int_x^\infty \frac{\exp(-y)}{y} \ln y \, dy \qquad (33)$$

Similarly we can write

$$\int_{t_{ws}}^{\infty} g_s(t_a) f(t_a) dt_a =$$

$$\left(\frac{M(k_B T)^2 \tau_s^2}{(\lambda_w M \tau_w)^2 \beta^2 V_0}\right) \exp\left[\frac{u_s k_B T}{\lambda_s \tau_s M V_0} \exp\left(\frac{-\gamma V_0 \lambda_w \beta \tau_{ws}}{k_B T}\right) \exp\left(\frac{\lambda_s}{k_B T} M V_0 t_{ws}\right)\right] \left[\begin{array}{l} G_1\left(\begin{array}{l}\frac{u_s k_B T}{\lambda_s \tau_s M V_0} \exp\left(\frac{-\gamma V_0 \lambda_w \beta \tau_{ws}}{k_B T}\right) \\ \exp\left(\frac{\lambda_s}{k_B T} M V_0 t_{ws}\right)\end{array}\right) \\ -\ln\left(\frac{u_s k_B T}{\lambda_s \tau_s M V_0}\right) \\ E_1\left(\begin{array}{l}\frac{u_s k_B T}{\lambda_s \tau_s M V_0} \exp\left(\frac{-\gamma V_0 \lambda_w \beta \tau_{ws}}{k_B T}\right) \\ \exp\left(\frac{\lambda_s}{kT} M V_0 t_{ws}\right)\end{array}\right) \end{array}\right]$$

(34)

Defining the non-dimensional parameters as following

$$V_w^* = \frac{kT}{\tau_w \lambda_w M}, \quad V_s^* = \frac{kT}{\tau_s \lambda_s M}, \quad \hat{V} = \frac{V}{V_w^*}, \quad \text{and} \quad \beta = \frac{V_w^*}{V_s^*} \qquad (35)$$

we can rewrite Eq 28, 29 and 31 as

$$g_w(t_{ws}) = \exp\left(-\frac{u_w}{\hat{V}_0} \exp(-\hat{\gamma}\hat{V}_0) \left\{\exp\left(\frac{\hat{V}_0 t_{ws}}{\tau_w}\right) - 1\right\}\right) \qquad (36)$$

$$g_s(t_a) = \exp\left(-\frac{u_s}{\beta \hat{V}_0} \exp(-\hat{\gamma}\hat{V}_0 \beta \tau_{ws}) \left\{\exp\left(\beta \hat{V}_0 \frac{t_a}{\tau_s}\right) - \exp\left(\beta \hat{V}_0 \frac{t_{ws}}{\tau_s}\right)\right\}\right) \qquad (37)$$

$$\frac{1}{\tau_w} \int_0^{t_{ws}} g_w(t_a) dt_a = \frac{1}{\hat{V}_0} \exp\left(\frac{u_w}{\hat{V}_0} \exp(-\hat{\gamma}\hat{V}_0)\right) \left\{E_1\left(\frac{u_w}{\hat{V}_0} \exp(-\hat{\gamma}\hat{V}_0)\right) - E_1\left(\frac{u_w}{\hat{V}_0} \exp(-\hat{\gamma}\hat{V}_0) \exp\left(\frac{\hat{V}_0 t_{ws}}{\tau_w}\right)\right)\right\}$$

(38)

$$\frac{1}{\tau_w} \int_{t_{ws}}^{\infty} g_s(t_a) dt_a = \frac{\tau_s}{\beta \hat{V}_0 \tau_w} \exp\left[\frac{u_s}{\beta \hat{V}_0} \exp(-\hat{\gamma}\hat{V}_0 \beta \tau_{ws}) \exp\left(\frac{\beta \hat{V}_0 t_{ws}}{\tau_s}\right)\right] E_1\left(\frac{u_s}{\beta \hat{V}_0} \exp(-\hat{\gamma}\hat{V}_0 \beta \tau_{ws}) \exp\left(\frac{\beta \hat{V}_0 t_{ws}}{\tau_s}\right)\right)$$

(39)

Substitution into Eq 26 gives

$$\frac{\sigma_0}{N_0} = \frac{\begin{bmatrix} \dfrac{k_B T}{\hat{V}_0 \lambda_w} \exp\left(\dfrac{u_w}{\hat{V}_0} e^{-\hat{\gamma}\hat{V}_0}\right) \begin{bmatrix} G_1\left(\dfrac{u_w}{\hat{V}_0} e^{-\hat{\gamma}\hat{V}_0}\right) - G_1\left(\dfrac{u_w}{\hat{V}_0} e^{-\hat{\gamma}\hat{V}_0} \exp\left(\hat{V}_0 \dfrac{t_{ws}}{\tau_w}\right)\right) - \\ \ln\left(\dfrac{u_w}{\hat{V}_0}\right) \left\{ \begin{array}{l} E_1\left(\dfrac{u_w}{\hat{V}_0} e^{-\hat{\gamma}\hat{V}_0}\right) - \\ E_1\left(\dfrac{u_w}{\hat{V}_0} e^{-\hat{\gamma}\hat{V}_0} \exp\left(\hat{V}_0 \dfrac{t_{ws}}{\tau_w}\right)\right) \end{array} \right\} \end{bmatrix} + \\ \dfrac{\tau_s k_B T}{\tau_w \lambda_s \beta \hat{V}_0} \exp\left[\dfrac{u_s}{\beta \hat{V}_0} \exp(-\hat{\gamma}\beta\hat{V}_0 \tau_{ws}) \exp\left(\dfrac{\beta\hat{V}_0 t_{ws}}{\tau_s}\right) - \dfrac{u_w}{\hat{V}_0} e^{-\hat{\gamma}\hat{V}_0} \left\{\exp\left(\dfrac{\hat{V}_0 t_{ws}}{\tau_w}\right) - 1\right\}\right] \\ \begin{bmatrix} G_1\left(\dfrac{u_s}{\beta\hat{V}_0} \exp(-\hat{\gamma}\beta\hat{V}_0 \tau_{ws}) \exp\left(\dfrac{\beta\hat{V}_0 t_{ws}}{\tau_s}\right)\right) \\ -\ln\left(\dfrac{u_s}{\beta\hat{V}_0}\right) E_1\left(\dfrac{u_s}{\beta\hat{V}_0} \exp(-\hat{\gamma}\beta\hat{V}_0 \tau_{ws}) \exp\left(\dfrac{\beta\hat{V}_0 t_{ws}}{\tau_s}\right)\right) \end{bmatrix} \\ +\gamma V_0 \begin{bmatrix} \dfrac{1}{\hat{V}_0} \exp\left(\dfrac{u_w}{\hat{V}_0} e^{-\hat{\gamma}\hat{V}_0}\right) \left\{ E_1\left(\dfrac{u_w}{\hat{V}_0} e^{-\hat{\gamma}\hat{V}_0}\right) - E_1\left(\dfrac{u_w}{\hat{V}_0} e^{-\hat{\gamma}\hat{V}_0} \exp\left(\dfrac{\hat{V}_0 t_{ws}}{\tau_w}\right)\right) \right\} \\ + \dfrac{\tau_s}{\beta\hat{V}_0 \tau_w} \exp\left[\dfrac{u_s}{\beta\hat{V}_0} \exp(-\hat{\gamma}\beta\hat{V}_0 \tau_{ws}) \exp\left(\dfrac{\beta\hat{V}_0 t_{ws}}{\tau_s}\right) - \dfrac{u_w}{\hat{V}_0} e^{-\hat{\gamma}\hat{V}_0} \left\{\exp\left(\dfrac{\hat{V}_0 t_{ws}}{\tau_w}\right) - 1\right\}\right] \\ E_1\left(\dfrac{u_s}{\beta\hat{V}_0} \exp(-\hat{\gamma}\beta\hat{V}_0 \tau_{ws}) \exp\left(\dfrac{\beta\hat{V}_0 t_{ws}}{\tau_s}\right)\right) \end{bmatrix} \end{bmatrix}}{1 + \begin{bmatrix} \dfrac{1}{\hat{V}_0} \exp\left(\dfrac{u_w}{\hat{V}_0} e^{-\hat{\gamma}\hat{V}_0}\right) \left\{ E_1\left(\dfrac{u_w}{\hat{V}_0} e^{-\hat{\gamma}\hat{V}_0}\right) - E_1\left(\dfrac{u_w}{\hat{V}_0} e^{-\hat{\gamma}\hat{V}_0} \exp\left(\dfrac{\hat{V}_0 t_{ws}}{\tau_w}\right)\right) \right\} \\ + \dfrac{\tau_s}{\beta\hat{V}_0 \tau_w} \exp\left[\dfrac{u_s}{\beta\hat{V}_0} \exp(-\hat{\gamma}\beta\hat{V}_0 \tau_{ws}) \exp\left(\dfrac{\beta\hat{V}_0 t_{ws}}{\tau_s}\right) - \dfrac{u_w}{\hat{V}_0} e^{-\hat{\gamma}\hat{V}_0} \left\{\exp\left(\dfrac{\hat{V}_0 t_{ws}}{\tau_w}\right) - 1\right\}\right] \\ E_1\left(\dfrac{u_s}{\beta\hat{V}_0} \exp(-\hat{\gamma}\beta\hat{V}_0 \tau_{ws}) \exp\left(\dfrac{\beta\hat{V}_0 t_{ws}}{\tau_s}\right)\right) \end{bmatrix}}$$

(40)

Defining

$$\sigma^* = \frac{k_B T N_0}{\lambda_w} \quad , \quad \hat{S} = \frac{S}{S^*} \quad \text{and} \quad \hat{\gamma} = \frac{\gamma}{\tau_w M} \tag{41}$$

We write

$$\hat{\sigma}_0 = \frac{\begin{bmatrix} \dfrac{1}{\hat{V}_0} \exp\left(\dfrac{u_w}{\hat{V}_0} e^{-\hat{\gamma}\hat{V}_0}\right) \begin{bmatrix} G_1\left(\dfrac{u_w}{\hat{V}_0} e^{-\hat{\gamma}\hat{V}_0}\right) - G_1\left(\dfrac{u_w}{\hat{V}_0} e^{-\hat{\gamma}\hat{V}_0} \exp\left(\hat{V}_0 \dfrac{t_{ws}}{\tau_w}\right)\right) - \\ \ln\left(\dfrac{u_w}{\hat{V}_0}\right)\left\{E_1\left(\dfrac{u_w}{\hat{V}_0} e^{-\hat{\gamma}\hat{V}_0}\right) - E_1\left(\dfrac{u_w}{\hat{V}_0} e^{-\hat{\gamma}\hat{V}_0} \exp\left(\hat{V}_0 \dfrac{t_{ws}}{\tau_w}\right)\right)\right\} \end{bmatrix} + \\ \dfrac{1}{\beta^2 \hat{V}_0 \tau_{ws}^2} \exp\left[\dfrac{u_s}{\beta \hat{V}_0} \exp(-\hat{\gamma}\hat{V}_0 \beta \tau_{ws}) \exp\left(\dfrac{\beta \hat{V}_0 t_{ws}}{\tau_s}\right) - \dfrac{u_w}{\hat{V}_0} e^{-\hat{\gamma}\hat{V}_0}\left\{\exp\left(\dfrac{\hat{V}_0 t_{ws}}{\tau_w}\right) - 1\right\}\right] \\ \begin{bmatrix} G_1\left(\dfrac{u_s}{\beta \hat{V}_0} \exp(-\hat{\gamma}\hat{V}_0 \beta \tau_{ws}) \exp\left(\dfrac{\beta \hat{V}_0 t_{ws}}{\tau_s}\right)\right) \\ -\ln\left(\dfrac{u_s}{\beta \hat{V}_0}\right) E_1\left(\dfrac{u_s}{\beta \hat{V}_0} \exp(-\hat{\gamma}\hat{V}_0 \beta \tau_{ws}) \exp\left(\dfrac{\beta \hat{V}_0 t_{ws}}{\tau_s}\right)\right) \end{bmatrix} \\ +\hat{\gamma}\begin{bmatrix} \exp\left(\dfrac{u_w}{\hat{V}_0} e^{-\hat{\gamma}\hat{V}_0}\right)\left\{E_1\left(\dfrac{u_w}{\hat{V}_0} e^{-\hat{\gamma}\hat{V}_0}\right) - E_1\left(\dfrac{u_w}{\hat{V}_0} e^{-\hat{\gamma}\hat{V}_0} \exp\left(\dfrac{\hat{V}_0 t_{ws}}{\tau_w}\right)\right)\right\} \\ + \dfrac{\tau_s}{\beta \tau_w} \exp\left[\dfrac{u_s}{\beta \hat{V}_0} \exp(-\hat{\gamma}\hat{V}_0 \beta \tau_{ws}) \exp\left(\dfrac{\beta \hat{V}_0 t_{ws}}{\tau_s}\right) - \dfrac{u_w}{\hat{V}_0} e^{-\hat{\gamma}\hat{V}_0}\left\{\exp\left(\dfrac{\hat{V}_0 t_{ws}}{\tau_w}\right) - 1\right\}\right] \\ E_1\left(\dfrac{u_s}{\beta \hat{V}_0} \exp(-\hat{\gamma}\hat{V}_0 \beta \tau_{ws}) \exp\left(\dfrac{\beta \hat{V}_0 t_{ws}}{\tau_s}\right)\right) \end{bmatrix} \end{bmatrix}}{1+\begin{bmatrix} \dfrac{1}{\hat{V}_0} \exp\left(\dfrac{u_w}{\hat{V}_0} e^{-\hat{\gamma}\hat{V}_0}\right)\left\{E_1\left(\dfrac{u_w}{\hat{V}_0} e^{-\hat{\gamma}\hat{V}_0}\right) - E_1\left(\dfrac{u_w}{\hat{V}_0} e^{-\hat{\gamma}\hat{V}_0} \exp\left(\dfrac{\hat{V}_0 t_{ws}}{\tau_w}\right)\right)\right\} \\ + \dfrac{\tau_s}{\beta \hat{V}_0 \tau_w} \exp\left[\dfrac{u_s}{\beta \hat{V}_0} \exp(-\hat{\gamma}\hat{V}_0 \beta \tau_{ws}) \exp\left(\dfrac{\beta \hat{V}_0 t_{ws}}{\tau_s}\right) - \dfrac{u_w}{\hat{V}_0} e^{-\hat{\gamma}\hat{V}_0}\left\{\exp\left(\dfrac{\hat{V}_0 t_{ws}}{\tau_w}\right) - 1\right\}\right] \\ E_1\left(\dfrac{u_s}{\beta \hat{V}_0} \exp(-\hat{\gamma}\hat{V}_0 \beta \tau_{ws}) \exp\left(\dfrac{\beta \hat{V}_0 t_{ws}}{\tau_s}\right)\right) \end{bmatrix}} \quad (42)$$

It can be shown that strong bond friction model reduces to weak bond friction model under the conditions: $\beta = V_w^*/V_s^* = 1$, $u_w/u_s = 1$, $\lambda_w/\lambda_s = 1$, $\tau_w/\tau_s = 1$ and $t_{ws} = 0$,

**Solution for relaxation process**

Here we modify Eq 42 by replacing $\hat{V}_0$ by $\hat{V}_c$ and $\hat{\sigma}_0$ by $\hat{\sigma}_c(t)$, the latter being a function of time.

$$\hat{\sigma}_c(t) = \frac{\begin{bmatrix} \frac{1}{\hat{V}_c}\exp\left(\frac{u_w}{\hat{V}_c}e^{-\hat{\gamma}\hat{V}_c}\right)\begin{bmatrix} G_1\left(\frac{u_w}{\hat{V}_c}e^{-\hat{\gamma}\hat{V}_c}\right) - G_1\left(\frac{u_w}{\hat{V}_c}e^{-\hat{\gamma}\hat{V}_c}\exp\left(\hat{V}_c\frac{t_{ws}}{\tau_w}\right)\right) - \\ \ln\left(\frac{u_w}{\hat{V}_c}\right)\begin{Bmatrix} E_1\left(\frac{u_w}{\hat{V}_c}e^{-\hat{\gamma}\hat{V}_c}\right) \\ -E_1\left(\frac{u_w}{\hat{V}_c}e^{-\hat{\gamma}\hat{V}_c}\exp\left(\hat{V}_c\frac{t_{ws}}{\tau_w}\right)\right) \end{Bmatrix} \end{bmatrix} + \\ \frac{1}{\beta^2\hat{V}_c\tau_{ws}^2}\exp\begin{bmatrix} \frac{u_s}{\beta\hat{V}_c}\exp(-\hat{\gamma}\hat{V}_c\beta\tau_{ws})\exp\left(\frac{\beta\hat{V}_ct_{ws}}{\tau_s}\right) \\ -\frac{u_w}{\hat{V}_c}e^{-\hat{\gamma}\hat{V}_c}\left\{\exp\left(\frac{\hat{V}_ct_{ws}}{\tau_w}\right)-1\right\} \end{bmatrix} \\ \begin{bmatrix} G_1\left(\frac{u_s}{\beta\hat{V}_c}\exp(-\hat{\gamma}\hat{V}_c\beta\tau_{ws})\exp\left(\frac{\beta\hat{V}_ct_{ws}}{\tau_s}\right)\right) \\ -\ln\left(\frac{u_s}{\beta\hat{V}_c}\right)E_1\left(\frac{u_s}{\beta\hat{V}_c}\exp(-\hat{\gamma}\hat{V}_c\beta\tau_{ws})\exp\left(\frac{\beta\hat{V}_ct_{ws}}{\tau_s}\right)\right) \end{bmatrix} \\ +\hat{\gamma}\begin{bmatrix} \exp\left(\frac{u_w}{\hat{V}_c}e^{-\hat{\gamma}\hat{V}_c}\right)\left\{E_1\left(\frac{u_w}{\hat{V}_c}e^{-\hat{\gamma}\hat{V}_c}\right) - E_1\left(\frac{u_w}{\hat{V}_c}e^{-\hat{\gamma}\hat{V}_c}\exp\left(\frac{\hat{V}_ct_{ws}}{\tau_w}\right)\right)\right\} \\ +\frac{\tau_s}{\tau_w\beta}\exp\begin{bmatrix} \frac{u_s}{\beta\hat{V}_c}\exp(-\hat{\gamma}\hat{V}_c\beta\tau_{ws})\exp\left(\frac{\beta\hat{V}_ct_{ws}}{\tau_s}\right) \\ -\frac{u_w}{\hat{V}_c}e^{-\hat{\gamma}\hat{V}_c}\left\{\exp\left(\frac{\hat{V}_ct_{ws}}{\tau_w}\right)-1\right\} \end{bmatrix} \\ E_1\left(\frac{u_s}{\beta\hat{V}_c}\exp(-\hat{\gamma}\hat{V}_c\beta\tau_{ws})\exp\left(\frac{\beta\hat{V}_ct_{ws}}{\tau_s}\right)\right) \end{bmatrix} \end{bmatrix}}{\begin{bmatrix} \frac{1}{\hat{V}_c}\exp\left(\frac{u_w}{\hat{V}_c}e^{-\hat{\gamma}\hat{V}_c}\right)\begin{Bmatrix} E_1\left(\frac{u_w}{\hat{V}_c}e^{-\hat{\gamma}\hat{V}_c}\right) - \\ E_1\left(\frac{u_w}{\hat{V}_c}e^{-\hat{\gamma}\hat{V}_c}\exp\left(\frac{\hat{V}_ct_{ws}}{\tau_w}\right)\right) \end{Bmatrix} + \\ 1+\frac{\tau_s}{\hat{V}_c\tau_w\beta}\exp\begin{bmatrix} \frac{u_s}{\beta\hat{V}_c}\exp(-\hat{\gamma}\hat{V}_c\beta\tau_{ws})\exp\left(\frac{\beta\hat{V}_ct_{ws}}{\tau_s}\right) \\ -\frac{u_w}{\hat{V}_c}e^{-\hat{\gamma}\hat{V}_c}\left\{\exp\left(\frac{\hat{V}_ct_{ws}}{\tau_w}\right)-1\right\} \end{bmatrix} \\ E_1\left(\frac{u_s}{\beta\hat{V}_c}\exp(-\hat{\gamma}\hat{V}_c\beta\tau_{ws})\exp\left(\frac{\beta\hat{V}_ct_{ws}}{\tau_s}\right)\right) \end{bmatrix}} \quad (43)$$

After converting non-dimensional form by assuming $u_{ws} = u_w/u_s$, $\tau_{ws} = \tau_w/\tau_s$, $t_{wsw} = t_{ws}/\tau_w$, $\lambda_{ws} = \lambda_w/\lambda_s$, $\beta = 1/\lambda_{ws}\tau_{ws}$ and $z = \frac{u_w}{\hat{V}_c} e^{-\hat{\gamma}\hat{V}_c}$.

$$\hat{\sigma}_c = \frac{\begin{bmatrix} \frac{1}{\hat{V}_c}\exp(z)\begin{bmatrix} G_1(z) \\ -G_1(z\exp(\hat{V}_c t_{wsw})) - \ln(ze^{\hat{\gamma}\hat{V}_c})\begin{Bmatrix} E_1(z) \\ -E_1(z\exp(\hat{V}_c t_{wsw})) \end{Bmatrix} \end{bmatrix} + \\ \frac{1}{\beta^2 \tau_{ws}^2 \hat{V}_c}\exp\begin{bmatrix} \frac{u_w}{\beta u_{ws}\hat{V}_c}\exp(-\hat{\gamma}\beta\tau_{ws}\hat{V}_c)\exp(\beta t_{wsw}\tau_{ws}\hat{V}_c) \\ -z\{\exp(\hat{V}_c t_{wsw})-1\} \end{bmatrix}\begin{bmatrix} G_1\left(\frac{u_w}{\beta u_{ws}\hat{V}_c}\exp(-\hat{\gamma}\beta\tau_{ws}\hat{V}_c)\exp(\beta t_{wsw}\tau_{ws}\hat{V}_c)\right) \\ -\ln\left(\frac{u_w}{\beta u_{ws}\hat{V}_c}\right) \\ E_1\left(\frac{u_w}{\beta u_{ws}\hat{V}_c}\exp(-\hat{\gamma}\beta\tau_{ws}\hat{V}_c)\exp(\beta t_{wsw}\tau_{ws}\hat{V}_c)\right) \end{bmatrix} \\ +\hat{\gamma}\begin{bmatrix} u_w \exp(z)\{E_1(z) - E_1(z\exp(\hat{V}_c t_{wsw}))\} \\ +\frac{u_w}{\tau_{ws}\beta}\exp\begin{bmatrix} \frac{u_s}{\beta\hat{V}_c}\exp(-\hat{\gamma}\beta\tau_{ws}\hat{V}_c)\exp(\hat{V}_c t_{wsw}\beta\tau_{ws}) \\ -z\{\exp(\hat{V}_c t_{wsw})-1\} \end{bmatrix} E_1\left(\frac{u_s}{\beta\hat{V}_c}\exp(-\hat{\gamma}\beta\tau_{ws}\hat{V}_c)\exp(\hat{V}_c t_{wsw}\beta\tau_{ws})\right) \end{bmatrix}}{u_w + \begin{bmatrix} \frac{1}{\hat{V}_c}\exp(z)\{E_1(z) - E_1(z\exp(\hat{V}_c t_{wsw}))\} \\ +\frac{1}{\tau_{ws}\beta\hat{V}_c}\exp\begin{bmatrix} \frac{u_s}{\beta\hat{V}_c}\exp(-\hat{\gamma}\beta\tau_{ws}\hat{V}_c)\exp(\hat{V}_c t_{wsw}\beta\tau_{ws}) \\ -z\{\exp(\hat{V}_c t_{wsw})-1\} \end{bmatrix} E_1\left(\frac{u_s}{\beta\hat{V}_c}\exp(-\hat{\gamma}\beta\tau_{ws}\hat{V}_c)\exp(\hat{V}_c t_{wsw}\beta\tau_{ws})\right) \end{bmatrix}}$$

(44)

Eq 44 should be solved in conjunction with Eq 2, i.e., $d\sigma_c(t)/dt = -K_b V_c(t)$ which can be written in dimensionless form as

$$\frac{d\hat{\sigma}_c(\hat{t})}{d\hat{t}} = -\hat{V}_c \qquad (45)$$

where $\hat{t}=t/t^*$ where $t^*=\sigma^*/K_b V^*$. The expression for relaxation is differentiated with relaxation for solving $\hat{V}_c(\hat{t})$ and then plugging $\hat{V}_c(\hat{t})$ into the expression of steady relaxation for getting $\hat{\sigma}_c(\hat{t})$.

Eqs 44 and 45 together form a set of differential-algebraic equations. Equations can be solved for $\sigma_c(t)$, using the following initial condition at $\hat{t}=0$ $\varepsilon=1$, $\hat{\sigma}_c=\hat{\sigma}_0$ where $\hat{\sigma}_0$ is the stress experienced by the block during the steady motion prior to stoppage.

**Parametric Study:**

The parametric study of the relaxation model was carried out to understand the effect of friction parameters on relaxation stress. The value of dimensionless friction parameters are realistically chosen as $\hat{V}_0=5$, $u_w=1, u_{ws}=1.5, t_{wsw}=2, \tau_{ws}=0.5, \lambda_{ws}=3$. An important parametric study is the effect of pulling velocity $\hat{V}_0$, on relaxation stress $\hat{\sigma}_c(\hat{t})$. Experimentally it is seen that although initial value of relaxation stress increases with $\hat{V}_0$, it ultimately settles to the same non-zero residual stress level[1]. The results are presented in Fig.1 for pulling velocity $\hat{V}_0 =$ 5.0, 9.0 and 14.0. It may be seen that irrespective of $\hat{V}_0$, it relaxes to the same level of residual stress. This observation is consistent with the experimental observations[1].

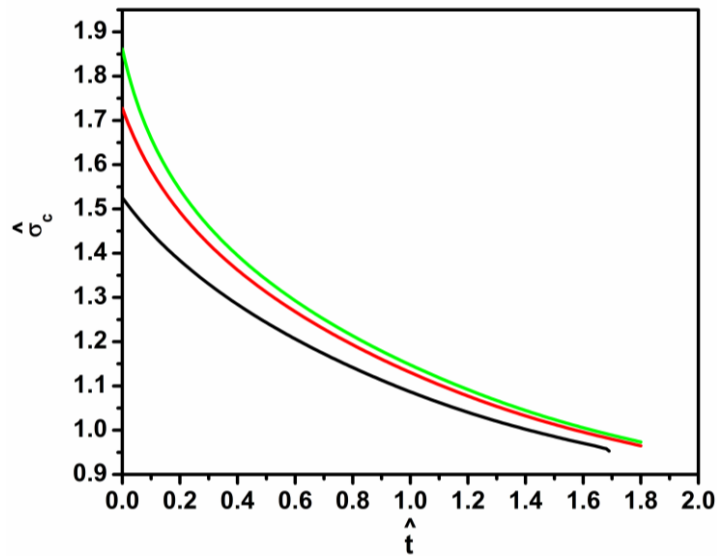

**Fig.1** Effect of pulling velocity $\hat{V}_0$ = 5.0 (black), 9.0 (red) and 14.0 (green) on residual stress vs. relaxation time process.

**Effect of $u_s$ us on relaxation stress:** The effect of $u_s$ on the relaxation process is studied in the simulation. It is assumed that $u_s$ is always less than $u_w$. It may be seen in Fig.2 that relaxation residual stress $\hat{\sigma}_c$ increases upon decrease of $u_s$ from $u_s = 0.66$ to $u_s = 0.1$. This because stronger bond forms with decreasing $u_s$.

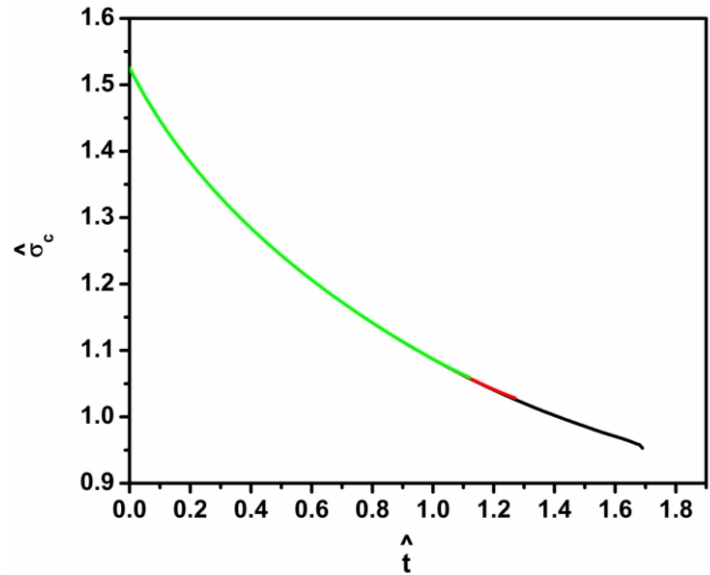

**Fig.2** Effect $u_s$ = 0.66 (black), 0.3(red) and 0.1 (green color) on time dependent relaxation stress $\hat{\sigma}_c(\hat{t})$.

**Effect of $t_{wsw}$ on relaxation stress:** This friction parameter basically controls the transition of weak bond to strong bond. It is seen in Fig.3 that $\hat{\sigma}_c(\hat{t})$ stabilizes to lower residual stress if $t_{wsw}$ changes from $t_{wsw}$ = 1 to 2. This observation is due to longer transition time $t_{wsw}$.

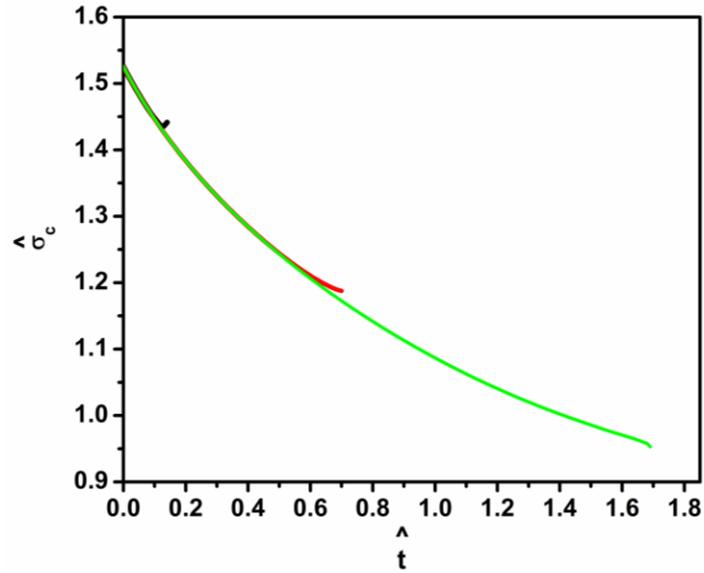

**Fig.3** Effect of transition time $t_{wsw}$ = 1(black), 1.5 (red), 2 (green) on relaxation stress vs. time.

**Effect of $\tau_{ws}$ on relaxation stress:**

Friction parameter $\tau_{ws} = \tau_w/\tau_s$ is related with the retraction time of a polymer chain between rupture and reformation of a bond. In other words, it is time between rupture and formation of a bond with the substrate. It is seen in Fig.4 that stabilization of $\hat{\sigma}_c(\hat{t})$ takes longer time upon increase of $\tau_{ws}$ =0.1, 0.3 and 0.5. The observation is owing to smaller value of $\tau_{ws}$ which results in stronger bonds thus the relaxation process becomes difficult. It is to be noted that for smaller value of $\tau_s$ results in larger value of $\tau_{ws}$ for a fixed $\tau_w$.

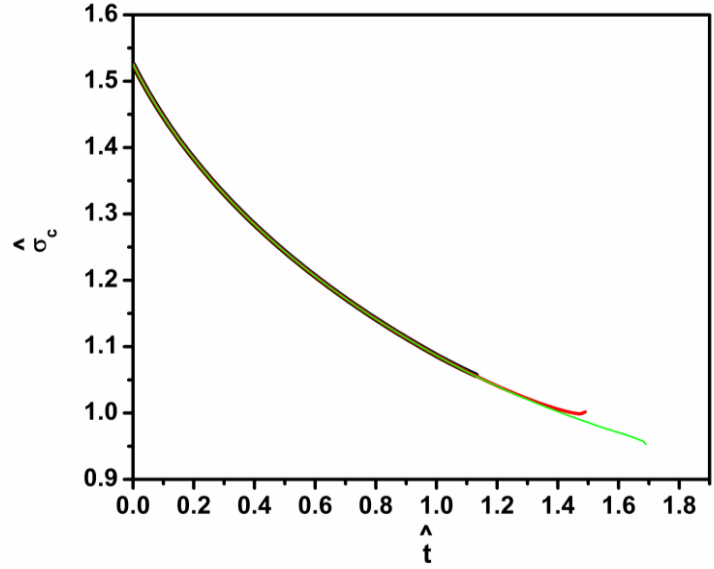

**Fig.4** Effect of strong bond friction parameter $\tau_{ws}$ = 0.1(black), 0.3(red) and 0.5 (green) on relaxation stress vs. time.

**Effect of $\lambda_{ws}$ on relaxation stress:**

The trend of $\lambda_{ws}$ on time dependent relaxation process is similar to $\tau_{ws}$. Friction parameter $\lambda_{ws} = \lambda_w/\lambda_s$ is basically the ratio of activation length between weak and strong bond. Since $\lambda_{ws}$ increases as $\lambda_{ws}$ = 1.0, 2.0 and 4.0 which means $\lambda_s$ decreases. For a smaller value of activation length $\lambda_s$ for a fixed $\lambda_w$ results in larger value of $\lambda_{ws}$. As mentioned earlier that a strong bond dampens more the relaxation process than the weak bond, thus it takes longer time to stabilize to residual stress. This observation is consistent with the simulation results in Fig.5

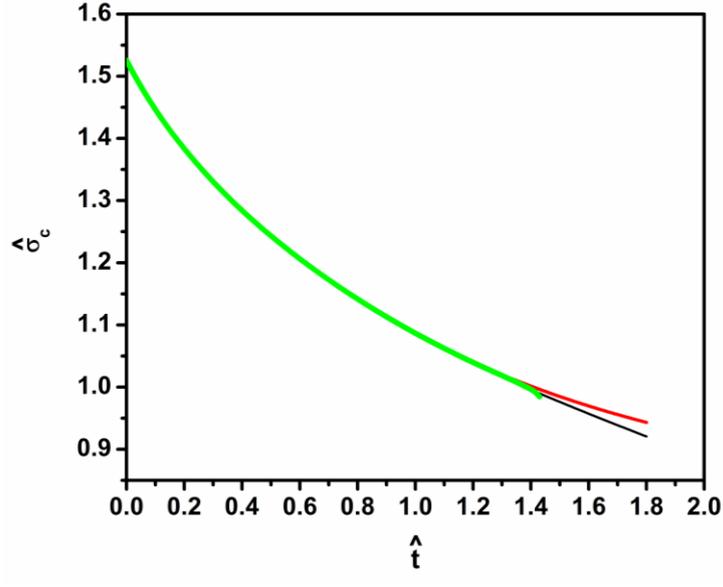

**Fig.5** Effect of friction parameter $\lambda_{ws}$ = 1.0(black), 2(red) and 4.0 (green) on time dependent relaxation stress.

Finally the weak bond model is also derived with the present relaxation model under the condition $u_{ws} = u_w = 1, t_{wss} = 0, \tau_{ws} = 1, \lambda_{ws} = 1$. It is seen in Fig6 that relaxation stress stabilizes to zero stress for the present case. And this is expected from the theory of weak bond model[5].

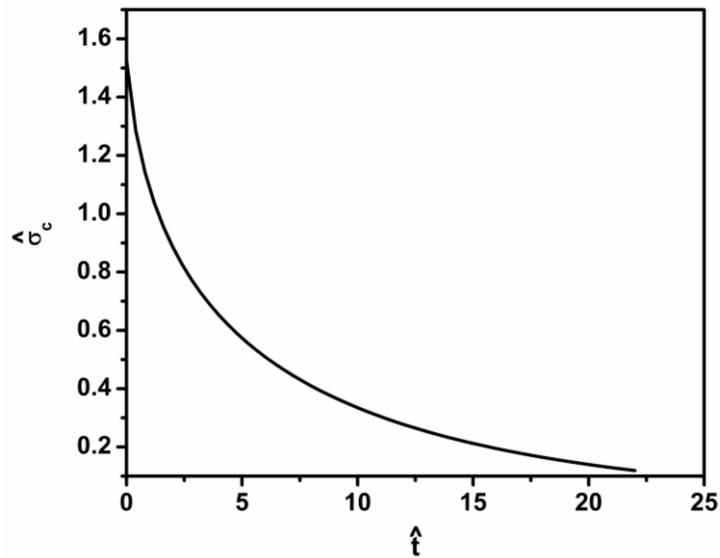

**Fig.6** Strong bond model under the condition $u_{ws} = u_w = 1$ for pulling velocity $\hat{V}_0 = 5$.

**Experimental validation:**

The present model is validated with the experimental data of Baumberger et al[1] as is seen in Fig.7. And the corresponding values of the parameters are given below as :

$\sigma^* = 1.77$ kPa, $V^* = 2.43 \times 10^{-4}$ ms$^{-1}$, $u_w = 1$, g=0.1895, $u_{ws} = 1.5$, $t_{wsw} = 2.85$, $\tau_{ws} = 1$, $\lambda_{ws} = 2.5$ and stiffness of the gel block is $K_b = 230.0$ kPa and pulling velocity of the gel block is $V_0 = 0.37$ mm s$^{-1}$. For weak bond, assuming that $\tau_w \sim 1$ microsecond, $\lambda_w \sim 1$ nm, it is then $\tau_s = 1$ microsecond, $\lambda_s = 0.4$ nm and $t_{ws} = 2.85$ microsecond. Using $\sigma^*$ and $V^*$, number of chains per unit area was obtained $n_0 \sim 10^{16}$ and stiffness of chain was $M \sim 0.1$ mNm$^{-1}$. These estimated values are in expected range.

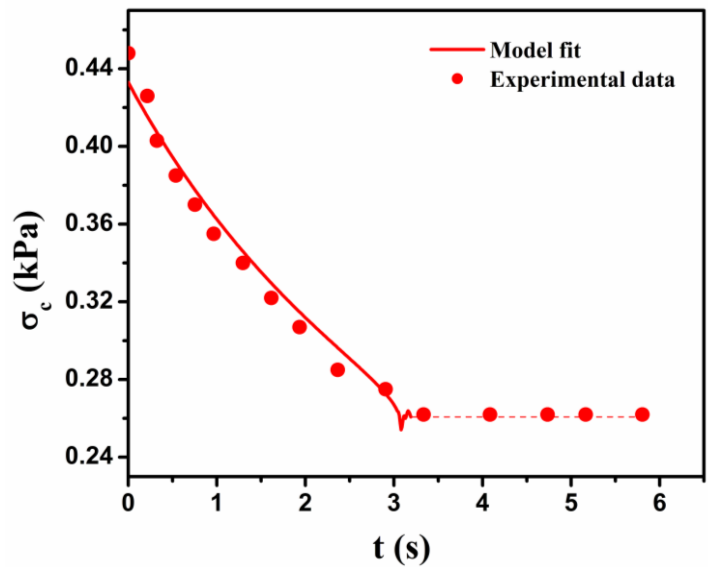

Fig.7 Curve fitting of strong bond model with relaxation data of Baumberger *et al.* (2003) for gelatin concentration, c=5%.

It will be interesting to validate the present strong bond model on other surfaces such as elastomers and rubbers for instance PDMS rubber. However, the Hookean law based strong bond model may not be valid in the case of PDMS rubbers. Therefore one has to modify considering the non-linear chain model such as FENE model. It is believed that the present understanding of frictional relaxation will also be useful for understanding static friction and stick-slip process.

**Conclusions:** A model for relaxation stress is proposed and validated with experimentally. Parametric studies concerning the model are also carried out in detailed.

Finally, it is also shown that this model reduces to weak bond model under the special condition. The relaxation model explains well the experimental observations.

**Appendix-1**

**Model-1**

1. Evaluation of $\int_0^{t_{ws}} g_w(t_a) dt_a$

$$\int_0^{t_{ws}} g_w(t_a) dt_a = \int_0^{t_{ws}} exp\left(-\frac{u_w kT}{\lambda_w \tau_w MV_0}\left\{exp\left(\frac{\lambda_w}{kT}MV_0 t_a\right)-1\right\}\right) dt_a$$

$$= exp\left(\frac{u_w kT}{\lambda_w \tau_w MV_0}\right) \int_0^{t_{ws}} exp\left(-\frac{u_w kT}{\lambda_w \tau_w MV_0} exp\left(\frac{\lambda_w}{kT}MV_0 t_a\right)\right) dt_a$$

Let $y = \frac{u_w kT}{\lambda_w \tau_w MV_0} exp\left(\frac{\lambda_w}{kT}MV_0 t_a\right)$

$$dt_a = \frac{dy}{\frac{u_w}{\tau_w}\left\{exp\left(\frac{\lambda_w}{kT}MV_0 t_a\right)\right\}} = \frac{kT dy}{y \lambda_w MV_0}$$

$$\int_0^{t_{ws}} g_w(t_a) dt_a = \frac{kT}{\lambda_w MV_0} exp\left(\frac{u_w kT}{\lambda_w \tau_w MV_0}\right) \int_{\frac{u_w kT}{\lambda_w \tau_w MV_0}}^{\frac{u_w kT}{\lambda_w \tau_w MV_0} exp\left(\frac{\lambda_w}{kT}MV_0 t_{ws}\right)} \frac{exp(-y) dy}{y}$$

We define

$$E_1(x) = \int_x^\infty \frac{e^{-y}}{y} dy$$

$$\int_0^{t_{ws}} g_w(t_a) dt_a = \frac{kT}{\lambda_w MV_0} exp\left(\frac{u_w kT}{\lambda_w \tau_w MV_0}\right) \left\{ E_1\left(\frac{u_w kT}{\lambda_w \tau_w MV_0}\right) - E_1\left(\frac{u_w kT}{\lambda_w \tau_w MV_0} exp\left(\frac{\lambda_w}{kT}MV_0 t_{ws}\right)\right) \right\}$$

Let us consider

2. Evaluation of $\int_{t_{ws}}^\infty g_s(t_a) dt_a$

$$g_s(t_a) = exp\left(-\frac{u_s kT}{\lambda_s \tau_s MV_0}\left\{ exp\left(\frac{\lambda_s}{kT}MV_0 t_a\right) - exp\left(\frac{\lambda_s}{kT}MV_0 t_{ws}\right)\right\}\right)$$

$$= exp\left[\frac{u_s kT}{\lambda_s \tau_s MV_0} exp\left(\frac{\lambda_s}{kT}MV_0 t_{ws}\right)\right] exp\left(-\frac{u_s kT}{\lambda_s \tau_s MV_0} exp\left(\frac{\lambda_s}{kT}MV_0 t_a\right)\right)$$

$$\int_{t_{ws}}^\infty g_s(t_a) dt_a = exp\left[\frac{u_s kT}{\lambda_s \tau_s MV_0} exp\left(\frac{\lambda_s}{kT}MV_0 t_{ws}\right)\right] \int_{t_{ws}}^\infty exp\left(-\frac{u_s kT}{\lambda_s \tau_s MV_0} exp\left(\frac{\lambda_s}{kT}MV_0 t_a\right)\right) dt_a$$

Let $y = \frac{u_s kT}{\lambda_s \tau_s MV_0} exp\left(\frac{\lambda_s}{kT}MV_0 t_a\right)$    $\frac{dy}{y} = \frac{\lambda_s}{kT}MV_0 dt_a$

$$\int_{t_{ws}}^\infty g_s(t_a) dt_a = \frac{kT}{\lambda_s MV_0} exp\left[\frac{u_s kT}{\lambda_s \tau_s MV_0} exp\left(\frac{\lambda_s}{kT}MV_0 t_{ws}\right)\right] \int_{\frac{u_s kT}{\lambda_s \tau_s MV_0} exp\left(\frac{\lambda_s}{kT}MV_0 t_{ws}\right)}^\infty \frac{exp(-y)}{y} dy$$

$$= \frac{kT}{\lambda_s MV_0} exp\left[\frac{u_s kT}{\lambda_s \tau_s MV_0} exp\left(\frac{\lambda_s}{kT}MV_0 t_{ws}\right)\right] E_1\left[\frac{u_s kT}{\lambda_s \tau_s MV_0} exp\left(\frac{\lambda_s}{kT}MV_0 t_{ws}\right)\right]$$

3. Evaluation of $\int_0^{t_{ws}} g_w(t_a) f(t_a) dt_a$

$$\int_0^{t_{ws}} g_w(t_a)f(t_a)dt_a = MV_0 \exp\left(\frac{u_w kT}{\lambda_w \tau_w MV_0}\right)\int_0^{t_{ws}} \exp\left(-\frac{u_w kT}{\lambda_w \tau_w MV_0}\exp\left(\frac{\lambda_w}{kT}MV_0 t_a\right)\right)t_a dt_a$$

Let $y = \dfrac{u_w kT}{\lambda_w \tau_w MV_0}\exp\left(\dfrac{\lambda_w}{kT}MV_0 t_a\right)$, $t_a = \dfrac{kT}{\lambda_w MV_0}\left[\ln y - \ln\left(\dfrac{u_w kT}{\lambda_w \tau_w MV_0}\right)\right]$

$$\int_0^{t_{ws}} g_w(t_a)f(t_a)dt_a = MV_0\left(\frac{kT}{\lambda_w MV_0}\right)^2 \exp\left(\frac{u_w kT}{\lambda_w \tau_w MV_0}\right) \int_{\frac{u_w kT}{\lambda_w \tau_w MV_0}}^{\frac{u_w kT}{\lambda_w \tau_w MV_0}\exp\left(\frac{\lambda_w}{kT}MV_0 t_{ws}\right)} \frac{\exp(-y)}{y}\left\{\begin{array}{l}\ln y \\ -\ln\left(\dfrac{u_w kT}{\lambda_w \tau_w MV_0}\right)\end{array}\right\} dy$$

$$= MV_0\left(\frac{kT}{\lambda_w MV_0}\right)^2 \exp\left(\frac{u_w kT}{\lambda_w \tau_w MV_0}\right)\left[\begin{array}{l}G_1\left(\dfrac{u_w kT}{\lambda_w \tau_w MV_0}\right) - G_1\left(\dfrac{u_w kT}{\lambda_w \tau_w MV_0}\exp\left(\dfrac{\lambda_w}{kT}MV_0 t_{ws}\right)\right) \\ -\ln\left(\dfrac{u_w kT}{\lambda_w \tau_w MV_0}\right)\left\{\begin{array}{l}E_1\left(\dfrac{u_w kT}{\lambda_w \tau_w MV_0}\right) \\ -E_1\left(\dfrac{u_w kT}{\lambda_w \tau_w MV_0}\exp\left(\dfrac{\lambda_w}{kT}MV_0 t_{ws}\right)\right)\end{array}\right\}\end{array}\right]$$

The integral $G_1(x)$ is defined as

$$G_1(x) = \int_x^\infty \frac{\exp(-y)}{y}\ln y\, dy$$

4. Evaluation of $\int_{t_{ws}}^\infty g_s(t_a)f(t_a)dt_a$

$$g_s(t_a) = \exp\left[\frac{u_s kT}{\lambda_s \tau_s MV_0}\exp\left(\frac{\lambda_s}{kT}MV_0 t_{ws}\right)\right]\exp\left(-\frac{u_s kT}{\lambda_s \tau_s MV_0}\exp\left(\frac{\lambda_s}{kT}MV_0 t_a\right)\right)$$

$$\int_0^{t_{ws}} g_s(t_a)f(t_a)dt_a$$
$$= MV_0 \exp\left[\frac{u_s kT}{\lambda_s \tau_s MV_0}\exp\left(\frac{\lambda_s}{kT}MV_0 t_{ws}\right)\right]\int_{t_{ws}}^\infty \exp\left(-\frac{u_s kT}{\lambda_s \tau_s MV_0}\exp\left(\frac{\lambda_s}{kT}MV_0 t_a\right)\right)t_a dt_a$$

Let $y = \dfrac{u_s kT}{\lambda_s \tau_s MV_0}\exp\left(\dfrac{\lambda_s}{kT}MV_0 t_a\right)$, $t_a = \dfrac{kT}{\lambda_s MV_0}\left[\ln y - \ln\left(\dfrac{u_s kT}{\lambda_s \tau_s MV_0}\right)\right]$, $\dfrac{dy}{y} = \dfrac{\lambda_s}{kT}MV_0 dt_a$

$$\int_{t_{ws}}^{\infty} g_s(t_a) f(t_a) dt_a = MV_0 \left( \frac{kT}{\lambda_s MV_0} \right)^2 exp\left[ \frac{u_s kT}{\lambda_s \tau_s MV_0} exp\left( \frac{\lambda_s}{kT} MV_0 t_{ws} \right) \right]$$

$$\int_{\frac{u_s kT}{\lambda_s \tau_s MV_0} exp\left( \frac{\lambda_s}{kT} MV_0 t_{ws} \right)}^{\infty} \frac{exp(-y)}{y} \left\{ \begin{array}{l} ln\, y \\ -ln\left( \frac{u_s kT}{\lambda_s \tau_s MV_0} \right) \end{array} \right\} dy$$

$$= MV_0 \left( \frac{kT}{\lambda_s MV_0} \right)^2 exp\left[ \frac{u_s kT}{\lambda_s \tau_s MV_0} exp\left( \frac{\lambda_s}{kT} MV_0 t_{ws} \right) \right] \left[ \begin{array}{l} G_1\left( \frac{u_s kT}{\lambda_s \tau_s MV_0} exp\left( \frac{\lambda_s}{kT} MV_0 t_{ws} \right) \right) \\ -ln\left( \frac{u_s kT}{\lambda_s \tau_s MV_0} \right) E_1\left( \frac{u_s kT}{\lambda_s \tau_s MV_0} exp\left( \frac{\lambda_s}{kT} MV_0 t_{ws} \right) \right) \end{array} \right]$$